\documentclass[aps,prl,twocolumn,showpacs,superscriptaddress]{revtex4}
\usepackage{graphicx}
\usepackage{textcomp}

\begin{document}

\title{Effects of Particle Shape on Growth Dynamics at Edges of Evaporating Colloidal Drops}

\author{Peter J. Yunker}
\affiliation{Department of Physics and Astronomy, University of Pennsylvania, Philadelphia PA 19104, USA}
\author{Matthew A. Lohr}
\affiliation{Department of Physics and Astronomy, University of Pennsylvania, Philadelphia PA 19104, USA}
\author{Tim Still}
\affiliation{Department of Physics and Astronomy, University of Pennsylvania, Philadelphia PA 19104, USA}
\affiliation{Complex Assemblies of Soft Matter, CNRS-Rhodia-University of Pennsylvania UMI 3254, Bristol, Pennsylvania 19007, USA}
\author{Alexei Borodin}
\affiliation{Department of Mathematics, Massachusetts Institute of Technology, Cambridge MA 02139, USA}
\author{D. J. Durian}
\affiliation{Department of Physics and Astronomy, University of Pennsylvania, Philadelphia PA 19104, USA}
\author{A. G. Yodh}
\affiliation{Department of Physics and Astronomy, University of Pennsylvania, Philadelphia PA 19104, USA}

\date{\today}
\begin{abstract}
We study the influence of particle shape on growth processes at the edges of evaporating drops. Aqueous suspensions of colloidal particles evaporate on glass slides, and convective flows during evaporation carry particles from drop center to drop edge, where they accumulate. The resulting particle deposits grow inhomogeneously from the edge in two-dimensions, and the deposition front, or growth line, varies spatio-temporally. Measurements of the fluctuations of the deposition front during evaporation enable us to identify distinct growth processes that depend strongly on particle shape. Sphere deposition exhibits a classic Poisson like growth process; deposition of slightly anisotropic particles, however, belongs to the Kardar-Parisi-Zhang (KPZ) universality class, and deposition of highly anisotropic ellipsoids appears to belong to a third universality class, characterized by KPZ fluctuations in the presence of quenched disorder.
\end{abstract}

\maketitle

Examples of surface and interfacial growth phenomena are diverse, ranging from the production of uniform coatings by vapor deposition of atoms onto a substrate \cite{PVD_review}, to burning paper wherein the combustion front roughens as it spreads \cite{KPZ_burning_paper, KPZ_burning_paper_II}, to bacterial colonies whose boundaries expand and fluctuate as bacteria replicate \cite{growth_process_bacteria}. The morphology of the resulting interfaces is a property that affects the macroscopic responses of such systems, and it is therefore desirable to relate interface morphology to the microscopic rules that govern growth \cite{family_vicsek_orig, growth_process_review, PVD_review}. To this end, simulations have directly compared a broad range of growth processes \cite{family_vicsek_orig,growth_process_review} and have found, for example, that the random deposition of repulsive particles is a Poisson process, while the random deposition of ``sticky'' particles belongs to a different universality class that leads to different interface morphology.

Besides discrete models, theoretical investigation of this problem has centered around continuum growth equations (e.g., \cite{EW_eq}).  One interesting approach that unified a large set of discrete simulations is based on the so-called Kardar-Parisi-Zhang (KPZ) equation \cite{KPZ_orig_PRL,growth_process_review,KPZ_review_II,KPZ_review_I,ht_dist_PRL,KPZ_2p1}.  This nonlinear equation relates stochastic growth and interfacial growth fronts/lines/surfaces to diffusion and local lateral correlations; its solutions are known and belong to the KPZ universality class \cite{KPZ_stationary_exact,KPZ_exact,KPZ_prob_dist,KPZ_freeenergy_fluc}.  The KPZ class presents a rare opportunity for connecting exact theoretical predictions about nonequilibrium growth phenomena with experiment. However, to date only a few members of the KPZ class have been experimentally identified \cite{KPZ_LC_PRL,KPZ_burning_paper,KPZ_burning_paper_II,growth_process_bacteria}. This paucity of KPZ examples is due, in part, to the presence of quenched disorder and long-range interactions in experiment, as well as to limited statistics, which make growth process differences difficult to discern.  In fact, broadly speaking, experimental confrontation of the microscopic rules explored by theory and simulation has been difficult.

In this contribution we demonstrate that the rich nonequilibrium physics of evaporating colloidal drops provides an attractive experimental system for study of such growth processes and for testing theory and simulation predictions. Specifically, we investigate the growth of particle deposits from the edges of evaporating colloidal drops. Particle deposition is observed by video microscopy at the single particle level.  Aqueous suspensions of colloidal particles are allowed to evaporate on glass slides at constant temperature and humidity, and radial convective flows during evaporation carry particles from drop center to drop edge, where they accumulate (Fig.\ 1a) \cite{coffeering_nagel_nat}. The resulting deposits of particles grow from the edge in two-dimensions, defining a deposition front, or growth line, that varies in space and time. Interestingly, these interfacial growth processes are strongly dependent on colloidal particle shape \cite{coffee_ring_ellipsoids}. Three distinct growth processes were discovered in the evaporating colloidal suspensions by tuning particle shape-dependent capillary interactions and thus varying the microscopic rules of deposition. The substantial shape fluctuations of the growth line of spheres are readily explained via a Poisson like deposition process; slightly anisotropic particles exhibit weaker fluctuations characteristic of KPZ class behavior, and very anisotropic ellipsoids exhibit behavior consistent with the KPZ class in the presence of quenched disorder \cite{KPZ_quencheddisorder,KPZ_quencheddis_SOC,KPZ_quencheddis_pinning}.

Our experiments employ water drops containing a suspension of micron-sized polystyrene spheres (Invitrogen) stretched asymmetrically to different aspect ratios \cite{ellipsoidstretch_PNAS,ellipsoid_stretch_first,coffee_ring_ellipsoids}. The spheres are $1.3$~\textmu m in diameter; all ellipsoids are stretched from these same $1.3$~\textmu m spheres. We evaporate the drops on glass slides (Fisher Scientific) and study suspensions containing particles of the same composition, but with different major-minor diameter aspect ratio ($\varepsilon$), including spheres ($\varepsilon=1.0$), slightly anisotropic particles ($\varepsilon=1.05,1.1,1.2$), and ellipsoids ($\varepsilon=1.5,2.5,3.5$). We study volume fractions ($\phi$) that vary from $\phi = 10^{-4}$ to $0.02$.

The experiments are reproducible across many droplets, enabling the accumulation of sufficient statistics to test continuum equation predictions of surface roughness scaling, and more.  Further, the strong shape-based capillary attractions between particles on the air-water interface \cite{ellipsoid_wetting_yodh,capillary_interactions_ellipsoids_yodh,interface_attract_plates,interface_anisotropicparticles,ellipsoids_interface_rheology,emulsions_ellipsoids_softmatter,interface_att_furst} permit  us to establish relationships between particle interaction and interfacial growth processes.

\begin{figure}
\scalebox{1.0}{\includegraphics{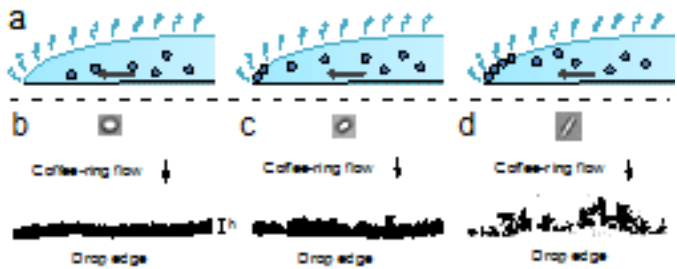}}
\caption{ a. Cartoon depicting the deposition mechanism. A radially outward flow carries particles from the drop center to the drop edge, where they are deposited on the air-water interface. b-d. Binarized experimental images of deposits of spheres ($\varepsilon=1.0$) (b), slightly stretched particles ($\varepsilon=1.2$) (c), and ellipsoids ($\varepsilon=2.5$) (d), along with images of single particles. A label in (b) demonstrates our definition of height ($h$), i.e., distance from drop edge. The drop edges and the direction of the coffee-ring driving flow are indicated. The probed window size $L$ is defined for $L=100$ \textmu m (b) and $L=25$ \textmu m (d). These scale bars also hold for (c). e-g. The deposit height profile (growth line), $h$, plotted as a function lateral position, $x$, at four different times, for $\varepsilon=1.0,1.2$ and $2.5$ (e-g, respectively). \label{fig:modes}}
\end{figure}

Qualitative differences between deposits crafted from particles with different aspect ratio, $\varepsilon$, are readily apparent (Fig.\ 1b-d). Deposits of spheres are densely packed (Fig.\ 1b); deposits of slightly anisotropic particles are loosely packed (Fig.\ 1c); deposits of very anisotropic ellipsoids form open networks and ``empty'' structures (Fig.\ 1d).   Deposits are characterized by their height, $h$, measured from the three-phase contact line (i.e., their penetration from the drop edge, measured radially inward) (see Fig.\ 1b). The deposit morphology is well described by the contour of its height profile (growth line). Almost all of the deposits described in this letter are monolayers of particles adsorbed on the air-water interface. The sole exceptions are the very largest sphere deposits ($\varepsilon=1.0$) discussed below.

\begin{figure}[htp]
\resizebox{\columnwidth}{!}{\includegraphics[]{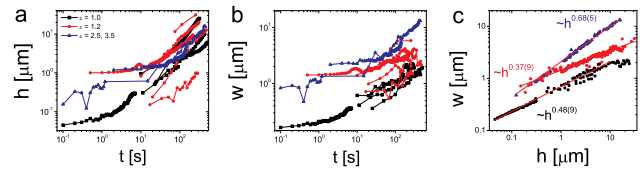}}
\caption{ Deposit width ($w$), i.e., the standard deviation of deposit height($h$), plotted versus $\bar{h}$ for spheres ($\varepsilon=1.0$, squares), slightly anisotropic particles ($\varepsilon=1.2$, circles), and ellipsoids ($\varepsilon=2.5,3.5$, triangles). Different colored dots indicate data derived from different experiments. The data collapse onto three unique trend lines based on $\varepsilon$. \label{fig:modes}}
\end{figure}

Deposit height varies spatially and temporally ($h(x,t)$). For example, the mean height, $\bar{h}$,  increases in time (see \cite{SOM_}). Spatial variation of $h$ is quantified by the standard deviation of $h$, a quantity referred to as the width, $w$. The growth of local fluctuations in $h$ produce an increase in $w$ over time \cite{SOM_}. Many growth processes exhibit a self-affine structure well-described by Family-Vicsek scaling. For example, $w\propto L^{\alpha}$ for small $L$, where $L$ is the horizontal (lateral) size of the window in which $w$ is calculated. Similarly, the width can exhibit power-law growth over time, i.e., $w \propto t^{\beta}$. Here $\alpha$ is called the roughness exponent and $\beta$ is called the growth exponent \cite{family_vicsek_orig,growth_process_review}. For simple Poisson processes, $\beta=1/2$, but $w$ does not depend on $L$, so $\alpha$ is poorly defined. For members of the KPZ universality class, $\beta=1/3$ and $\alpha=1/2$.

In exploring growth exponents, the use of real time, $t$, presented us with technical problems.   During evaporation, for example, the radially outward flow rate increases over time \cite{coffeering_nagel_nat}, preventing a controlled power law measurement. Further, the start time ($t=0$) is ambiguously defined; it could be when the drop is placed on the substrate, when the drop stops spreading, or when the first particle is deposited. Finally, the time-range during which deposits can form, i.e., before strong surface flows or particle aggregates artificially modify $w$, is often not long enough to extract meaningful power laws. By contrast, the use of average deposite height, $\bar{h}$, in place of time $t$ averts many of these technical issues.  $\bar{h}$ is linear with $t$ in all experiments, with a coefficient of determination $R^{2}>0.9$ in all cases. Importantly, use of $\bar{h}$ skirts the issue of increasing/decreasing flow rates, resolves ambiguities in defining $t=0$ ($\bar{h}=0$ is well defined), and enables us to perform many measurements with different drops, or on different parts of the same drop, all of which can be combined (Fig.\ 2c).  Thus, we are able to obtain sufficient statistics to provide data over two orders of magnitude in $\bar{h}$.

Three qualitatively and quantitatively different growth rate regimes are readily identifiable in the results (Fig.\ 2). For spheres ($\varepsilon=1.0$) $w \propto \bar{h}^{0.48(4)}$; for slightly anisotropic particles ($\varepsilon=1.2$) $w \propto \bar{h}^{0.37(4)}$; for ellipsoids with $\varepsilon = 2.5$ and $\varepsilon=3.5$, the data collapse onto a single curve with $w \propto \bar{h}^{0.68(4)}$ (the final digit uncertainty represents the standard error of the fit combined with measurement uncertainty). Three distinct growth processes are readily apparent. Spheres pack densely at the drop edge, but their growth deposit appears spatially uncorrelated; thus we observe a significant increase in $w$ over two decades in $\bar{h}$. Slightly anisotropic particles start with a relatively large $w$, which then increases slowly with $\bar{h}$. For the anisotropic ellipsoids, $w$ increases rapidly with $\bar{h}$, as dense regions grow at the expense of sparse regions, which remain sparse.

The measured growth exponent for slightly anisotropic particles ($\varepsilon=1.2$), $\beta=0.37(4)$, is consistent with the KPZ universality class, which predicts $\beta=1/3$ and $\alpha=1/2$.  We measure $\alpha$ based on the finite size scaling of $w$, as $w$ depends on the observation lengthscale, $L$. For small values of $L$, $w\propto L^{\alpha}$. (Note, at larger $L$, $w$ saturates or crosses over to a weaker power-law dependence.) The power-law dependence is measured over two decades in $L$ within the small $L$ regime (Fig.\ 3a). The best fit yields $\alpha=0.51(5)$. Both scaling exponents, $\beta=0.37(4)$ and $\alpha=0.51(5)$, are within experimental uncertainty of predicted values for the KPZ universality class.

\begin{figure}[htp]
\resizebox{\columnwidth}{!}{\includegraphics[]{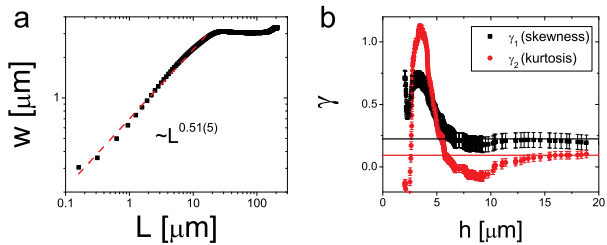}}
\caption{ a. Width, $w$, plotted versus probed lengthscale, $L$, for a drop containing slightly anisotropic particles $\varepsilon=1.2$ with height $\bar{h}=13$ ~\textmu m. The dashed line represents the best power law fit for $L<15$ \textmu m. The observed power law scaling is consistent with the KPZ universality class. b. Skewness ($\gamma_{1}$) and kurtosis ($\gamma_{2}$) of the $h$ distribution for particles with anisotropy $\varepsilon=1.2$ is plotted versus $h$. Solid lines represent values of $\gamma_{1}$ and $\gamma_{2}$ for Gaussian unitary ensemble matrices. Thus, the skewness and kurtosis imply that the distribution of height is consistent with the KPZ universality class. \label{fig:modes}}
\end{figure}

Next, we characterize the distribution of $h$, in samples containing particles with anisotropy $\varepsilon=1.2$. In particular,  the skewness, $\gamma_{1}(\bar{h}) = (1/L \int_{x=0..L} ((h(x,\bar{h})-\bar{h})^{3/2}) / w(\bar{h})^{3}) dx$, and excess kurtosis $\gamma_{2}(\bar{h}) = (1/L \int_{x=0..L} ((h(x,\bar{h})-\bar{h})^{4}) / w(\bar{h})^{2}) dx - 3$, are calculated; here $h(x,\bar{h})$ is the height at position $x$, at a time with mean height $\bar{h}$. For members of the KPZ class, the distribution of $h$ depends on the shape of the interface \cite{ht_dist_PRL}. For curved interfaces (e.g., the circular three-phase contact line of a sessile drop), a Gaussian unitary ensemble (GUE) random matrix distribution is expected ($\gamma_{1}=0.22$ and $\gamma_{2}=0.09$), while for flat interfaces, a Gaussian orthogonal ensemble distribution is expected ($\gamma_{1}=0.29$ and $\gamma_{2}=0.16$) \cite{ht_dist_PRL}. In our experiments, the height distribution levels out at $h \approx 15$ ~\textmu m where $\gamma_{1}=0.20(6)$ and $\gamma_{2}=0.10(3)$ (Fig.\ 3b). Thus, the distribution of height is consistent with predictions for a KPZ process with curved interface.

Finally, we note that particles with anisotropy $\varepsilon=1.2$ adsorb on the air-water interface and slightly deform the air-water interface \cite{ellipsoid_wetting_yodh}. These deformations, in turn, induce a relatively weak capillary interparticle attraction.  The attraction is significant at short-range, enabling particles to ``stick'' to each other once they reach the air-water interface. This behavior is similar to simulations of so-called ``ballistic deposition'' \cite{growth_process_review} which gives rise to KPZ dynamics.

For spheres, the measured growth exponent $\beta=0.48(4)$ (for $h<10$ ~\textmu m) is consistent with a Poisson process (raining particles) which might be expected, since spheres do not  significantly deform the air-water interface, and thus do not induce the strong quadrupolar attraction observed for ellipsoids. In this case, the distribution of heights is well fit by the Poisson distribution, $N(h) = \lambda^h exp(-\lambda) / h!$, where $N(h)$ is the number of occurrences of $h$, and $\lambda$ is the mean of the distribution. The best fit $\lambda$ increases linearly with $\bar{h}$ ($R^{2}=0.99$) \cite{SOM_}. The standard deviation of a Poisson distribution is $\sqrt{\bar{h}}$, and $\gamma_{1} = \bar{h}^{-1/2}$, i.e., the skewness of the height distribution decreases as the width increases. In our experiments, $\gamma_{1}$ does not approach its asymptotic values until $\bar{h}>1.0$~\textmu m, so a power law fit could only be measured over one decade. Instead, to determine whether these data are consistent with Poisson distributions, we plot $\gamma_{1} \cdot w$ (Fig.\ 4a). Since $w \propto \bar{h}^{0.5}$, we expect $\gamma_{1} \cdot w$ to be a constant when $\gamma_{1} \propto w^{-1} \approx \bar{h}^{-0.5}$. Experimentally, for $\varepsilon=1.0$, $\gamma_{1} \propto  \bar{h}^{-0.5}$, consistent with a Poisson process, and for $\varepsilon>1.0$, $\gamma_{1} \cdot  w$ is not consistent with a Poisson process.

We next characterized the lateral correlations in $w(\bar{h})$ for spheres. Unlike slightly anisotropic particles, spheres do not exhibit a distinct roughness exponent $\alpha$ \cite{SOM_}. The apparent absence of a characteristic roughness implies there is little spatial correlation in $h$, i.e., regions with large $h$ may directly neighbor regions with small $h$, which is again consistent with the Poisson distribution \cite{SOM_}. We note, however, that when the deposit is large (i.e., $h>10$ ~\textmu m), the deposit becomes multilayered and in three dimensions, more contacts are required to stabilize a particle, than in two dimensions. For $h>10$ ~\textmu m, the spheres are therefore able to find a local minimum in the height profile, rather than attaching to the first pair of particles they touch. The growth mechanism ceases to be a Poisson like process of random deposition at this point, and it is better described as random deposition with surface diffusion \cite{growth_process_review}.

For the very anisotropic ellipsoids, the measured growth exponent, $\beta=0.68(5)$, is consistent with a KPZ process in the presence of quenched disorder (KPZQ) \cite{KPZ_quencheddisorder,KPZ_quencheddis_pinning,KPZ_qd_anisotropicdepinning,KPZ_quencheddis_SOC}. Previous work found that if quenched disorder prevents interfacial growth in a particular region, then a new universality class (the KPZQ class) was produced with $\beta=0.68$ and $\alpha=0.63$ \cite{KPZ_quencheddisorder}. The measured roughness exponent for our experiments is $\alpha=0.61(2)$ (Fig.\ 4b). Thus the measured values of $\beta=0.68(5)$ and $\alpha=0.61(2)$ are consistent with the KPZQ class.

The latter observation was somewhat unexpected for us, because most KPZQ models are characterized by regions where growth is prevented, mixed with regions where growth is allowed \cite{KPZ_quencheddisorder,KPZ_quencheddis_pinning,KPZ_qd_anisotropicdepinning,KPZ_quencheddis_SOC}. Superficially, our experiments appear to have different conditions than those needed for KPZQ. However, the highly anisotropic ellipsoids induce strong capillary attraction on the air-water interface \cite{ellipsoid_wetting_yodh,capillary_interactions_ellipsoids_yodh,interface_attract_plates,interface_anisotropicparticles,ellipsoids_interface_rheology,emulsions_ellipsoids_softmatter,interface_att_furst}, which causes regions with many particles to strongly attract additional particles. Even particles that adsorb on the air-water interface in regions nearly void of particles are strongly attracted to particle rich regions, and eventually are deposited in particle rich regions. Thus, ellipsoids exhibit a colloidal ``Matthew Effect'' \cite{mattheweffect}. Surprisingly, this process is quite similar to the KPZQ scenario. The strong long-ranged capillary attraction enhances growth in particle rich regions, which effectively prevents (or at least slows) growth in other regions (Fig.\ 4c).

\begin{figure}
\scalebox{1.0}{\includegraphics{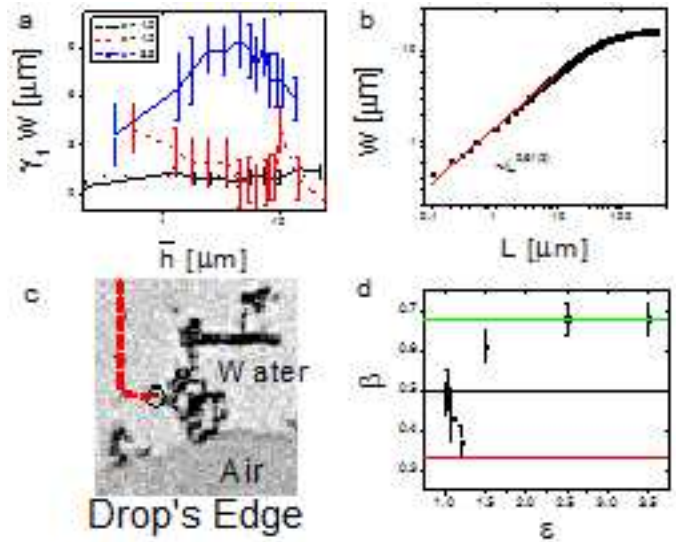}}
\caption{ a. Skewness, $\gamma_{1}$, multiplied by width, $w$, plotted versus $\bar{h}$ for $\varepsilon=1.0,1.2,3.5$ (solid line, dashed line, dotted line, respectively). The line with dashes and dots is $\gamma_{1} w=1$, the prediction for a Poisson process, i.e., the skewness grows linearly with the width. Error bars represent standard deviation. b. Width, $w$, plotted versus probed lengthscale, $L$, for a drop containing ellipsoids $\alpha=2.5$ with height $h=17$ ~\textmu m. c. Example of the colloidal ``Matthew Effect.'' The trajectory of an ellipsoid ($\varepsilon=2.5$) during deposition is shown (red line). Initially, it is pushed towards a region that contains very few ellipsoids. However, when it adsorbs on the air-water interface, it is attracted to a particle-rich region, and deposited there. d. Dynamic scaling exponent $\beta$ plotted versus particle aspect ratio $\varepsilon$. Three distinct regimes are readily identified (indicated by shading). Predictions for random deposition ($\beta=0.5$), KPZ processes ($\beta=1/3$), and KPZQ processes ($\beta=0.68$) are indicated by solid horizontal lines. \label{fig:modes}}
\end{figure}

Finally, we explored a range of aspect ratios which fall between these exemplary cases. The dynamic scaling exponent $\beta$ is shown in Fig.\ 4d for many different values of $\varepsilon$. Three regimes are readily identifiable. The dynamic scaling exponent $\beta$ starts at $\sim0.5$ for spheres. As $\varepsilon$ increases, $\beta$ abruptly decreases to $\sim0.35$. When $\varepsilon$ is increased further, $\beta$ increases to $\sim0.68$.

While quantitative differences between Poisson, KPZ, and KPZQ processes are provided by $\alpha$ and $\beta$, qualitative differences can be summarized by performing a series of simple simulations \cite{SOM_}. Consider a system wherein particles ``rain down'' vertically onto a 1D surface divided into a series of columns (imagine Tetris with individual blocks). A Poisson process can be modeled by randomly adding new particles without spatial or temporal correlation, i.e., the growth of each column is random and independent of neighboring columns. As a result, the tallest column may be next to the shortest column, and $w$ does not systematically depend on probed lengthscale. KPZ processes can be modeled following the same rules that govern Poisson processes, but utilizing ``sticky'' blocks which attach to the first particle they touch. Particles falling in short columns can then stick to the side of a tall column. This permits short columns that neighbor tall columns to grow very quickly; thus, $w$ grows slower in a KPZ process than in a Poisson process. Finally, KPZQ processes can be modeled following the same rules that govern KPZ processes, but with spatial modifications to the growth rate. Growth rates are assigned without spatial correlation such that some regions grow very quickly while other regions grow very slowly. This disparity in growth rates causes $w$ to increase very quickly.

To summarize, the growth process of particles deposited at the edge of evaporating suspensions is highly dependent on particle shape. Slightly anisotropic particles appear to be deposited by a KPZ process. Spheres are deposited by a Poisson like process, until the deposit becomes multilayered; at this point the process is best characterized as random deposition with surface diffusion. Finally, very anisotropic ellipsoids induce strong capillary attraction on the air-water interface and appear to belong to the KPZQ class, i.e., KPZ class with quenched disorder. Thus, evaporating drops of suspensions provide an appealing model system for studying theoretical predictions for different growth processes. Interesting future experiments could probe the aspect ratio boundary regions more comprehensively. Further, the various growth processes hold potentially important consequences for applications involving interfacial colloidal aggregates ( e.g., Pickering emulsions \cite{emulsions_ellipsoids_softmatter} and food processing \cite{food_emulsions,coffee_ring_ellipsoids_newsviews}), and it should be interesting to investigate how growth processes affect the stability or rigidity of colloidal deposits.

\begin{acknowledgments}
We thank Tom C. Lubensky for helpful discussions, and Barry Simon for bringing our experiments to A.B.'s attention. We gratefully acknowledge financial support from the National Science Foundation through DMR-0804881, the PENN MRSEC DMR11-20901, and NASA NNX08AO0G. A.B. gratefully acknowledges financial support from NSF grant DMS-1056390. T.S. acknowledges support from DAAD.
\end{acknowledgments}

\bibliographystyle{apsrev}

\begin{thebibliography}{36}
\expandafter\ifx\csname natexlab\endcsname\relax\def\natexlab#1{#1}\fi
\expandafter\ifx\csname bibnamefont\endcsname\relax
  \def\bibnamefont#1{#1}\fi
\expandafter\ifx\csname bibfnamefont\endcsname\relax
  \def\bibfnamefont#1{#1}\fi
\expandafter\ifx\csname citenamefont\endcsname\relax
  \def\citenamefont#1{#1}\fi
\expandafter\ifx\csname url\endcsname\relax
  \def\url#1{\texttt{#1}}\fi
\expandafter\ifx\csname urlprefix\endcsname\relax\def\urlprefix{URL }\fi
\providecommand{\bibinfo}[2]{#2}
\providecommand{\eprint}[2][]{\url{#2}}

\bibitem[{\citenamefont{Reichelt and Jiang}(1990)}]{PVD_review}
\bibinfo{author}{\bibfnamefont{K.}~\bibnamefont{Reichelt}} \bibnamefont{and}
  \bibinfo{author}{\bibfnamefont{X.}~\bibnamefont{Jiang}},
  \bibinfo{journal}{Thin Solid Films} \textbf{\bibinfo{volume}{191}},
  \bibinfo{pages}{91} (\bibinfo{year}{1990}), ISSN \bibinfo{issn}{00406090},
  \urlprefix\url{http://dx.doi.org/10.1016/0040-6090(90)90277-K}.

\bibitem[{\citenamefont{Maunuksela et~al.}(1997)\citenamefont{Maunuksela,
  Myllys, K\"{a}hk\"{o}nen, Timonen, Provatas, Alava, and
  Nissila}}]{KPZ_burning_paper}
\bibinfo{author}{\bibfnamefont{J.}~\bibnamefont{Maunuksela}},
  \bibinfo{author}{\bibfnamefont{M.}~\bibnamefont{Myllys}},
  \bibinfo{author}{\bibfnamefont{O.~P.} \bibnamefont{K\"{a}hk\"{o}nen}},
  \bibinfo{author}{\bibfnamefont{J.}~\bibnamefont{Timonen}},
  \bibinfo{author}{\bibfnamefont{N.}~\bibnamefont{Provatas}},
  \bibinfo{author}{\bibfnamefont{M.~J.} \bibnamefont{Alava}}, \bibnamefont{and}
  \bibinfo{author}{\bibfnamefont{T.~A.} \bibnamefont{Nissila}},
  \bibinfo{journal}{Physical Review Letters} \textbf{\bibinfo{volume}{79}},
  \bibinfo{pages}{1515} (\bibinfo{year}{1997}),
  \urlprefix\url{http://dx.doi.org/10.1103/PhysRevLett.79.1515}.

\bibitem[{\citenamefont{Myllys et~al.}(2001)\citenamefont{Myllys, Maunuksela,
  Alava, Nissila, Merikoski, and Timonen}}]{KPZ_burning_paper_II}
\bibinfo{author}{\bibfnamefont{M.}~\bibnamefont{Myllys}},
  \bibinfo{author}{\bibfnamefont{J.}~\bibnamefont{Maunuksela}},
  \bibinfo{author}{\bibfnamefont{M.}~\bibnamefont{Alava}},
  \bibinfo{author}{\bibfnamefont{T.~A.} \bibnamefont{Nissila}},
  \bibinfo{author}{\bibfnamefont{J.}~\bibnamefont{Merikoski}},
  \bibnamefont{and} \bibinfo{author}{\bibfnamefont{J.}~\bibnamefont{Timonen}},
  \bibinfo{journal}{Physical Review E} \textbf{\bibinfo{volume}{64}},
  \bibinfo{pages}{036101+} (\bibinfo{year}{2001}),
  \urlprefix\url{http://dx.doi.org/10.1103/PhysRevE.64.036101}.

\bibitem[{\citenamefont{Wakita et~al.}(1997)\citenamefont{Wakita, Itoh,
  Matsuyama, and Matsushita}}]{growth_process_bacteria}
\bibinfo{author}{\bibfnamefont{J.~I.} \bibnamefont{Wakita}},
  \bibinfo{author}{\bibfnamefont{H.}~\bibnamefont{Itoh}},
  \bibinfo{author}{\bibfnamefont{T.}~\bibnamefont{Matsuyama}},
  \bibnamefont{and}
  \bibinfo{author}{\bibfnamefont{M.}~\bibnamefont{Matsushita}},
  \bibinfo{journal}{Journal of the Physical Society of Japan}
  \textbf{\bibinfo{volume}{66}}, \bibinfo{pages}{67} (\bibinfo{year}{1997}),
  \urlprefix\url{http://dx.doi.org/10.1143/JPSJ.66.67}.

\bibitem[{\citenamefont{Family and Vicsek}(1985)}]{family_vicsek_orig}
\bibinfo{author}{\bibfnamefont{F.}~\bibnamefont{Family}} \bibnamefont{and}
  \bibinfo{author}{\bibfnamefont{T.}~\bibnamefont{Vicsek}},
  \bibinfo{journal}{Journal of Physics A: Mathematical and General}
  \textbf{\bibinfo{volume}{18}}, \bibinfo{pages}{L75} (\bibinfo{year}{1985}),
  ISSN \bibinfo{issn}{0305-4470},
  \urlprefix\url{http://dx.doi.org/10.1088/0305-4470/18/2/005}.

\bibitem[{\citenamefont{Family}(1990)}]{growth_process_review}
\bibinfo{author}{\bibfnamefont{F.}~\bibnamefont{Family}},
  \bibinfo{journal}{Physica A: Statistical Mechanics and its Applications}
  \textbf{\bibinfo{volume}{168}}, \bibinfo{pages}{561} (\bibinfo{year}{1990}),
  ISSN \bibinfo{issn}{03784371},
  \urlprefix\url{http://dx.doi.org/10.1016/0378-4371(90)90409-L}.

\bibitem[{\citenamefont{{Edwards} and {Wilkinson}}(1982)}]{EW_eq}
\bibinfo{author}{\bibfnamefont{S.~F.} \bibnamefont{{Edwards}}}
  \bibnamefont{and} \bibinfo{author}{\bibfnamefont{D.~R.}
  \bibnamefont{{Wilkinson}}}, \bibinfo{journal}{Royal Society of London
  Proceedings Series A} \textbf{\bibinfo{volume}{381}}, \bibinfo{pages}{17}
  (\bibinfo{year}{1982}),
  \urlprefix\url{http://dx.doi.org/10.1098/rspa.1982.0056}.

\bibitem[{\citenamefont{Kardar et~al.}(1986)\citenamefont{Kardar, Parisi, and
  Zhang}}]{KPZ_orig_PRL}
\bibinfo{author}{\bibfnamefont{M.}~\bibnamefont{Kardar}},
  \bibinfo{author}{\bibfnamefont{G.}~\bibnamefont{Parisi}}, \bibnamefont{and}
  \bibinfo{author}{\bibfnamefont{Y.~C.} \bibnamefont{Zhang}},
  \bibinfo{journal}{Physical Review Letters} \textbf{\bibinfo{volume}{56}},
  \bibinfo{pages}{889} (\bibinfo{year}{1986}),
  \urlprefix\url{http://dx.doi.org/10.1103/PhysRevLett.56.889}.

\bibitem[{\citenamefont{Kriecherbauer and Krug}(2010)}]{KPZ_review_II}
\bibinfo{author}{\bibfnamefont{T.}~\bibnamefont{Kriecherbauer}}
  \bibnamefont{and} \bibinfo{author}{\bibfnamefont{J.}~\bibnamefont{Krug}},
  \bibinfo{journal}{Journal of Physics A: Mathematical and Theoretical}
  \textbf{\bibinfo{volume}{43}}, \bibinfo{pages}{403001+}
  (\bibinfo{year}{2010}), ISSN \bibinfo{issn}{1751-8113},
  \urlprefix\url{http://dx.doi.org/10.1088/1751-8113/43/40/403001}.

\bibitem[{\citenamefont{Corwin}(2011)}]{KPZ_review_I}
\bibinfo{author}{\bibfnamefont{I.}~\bibnamefont{Corwin}}
  (\bibinfo{year}{2011}), \eprint{1106.1596},
  \urlprefix\url{http://arxiv.org/abs/1106.1596}.

\bibitem[{\citenamefont{Pr\"{a}hofer and Spohn}(2000)}]{ht_dist_PRL}
\bibinfo{author}{\bibfnamefont{M.}~\bibnamefont{Pr\"{a}hofer}}
  \bibnamefont{and} \bibinfo{author}{\bibfnamefont{H.}~\bibnamefont{Spohn}},
  \bibinfo{journal}{Physical Review Letters} \textbf{\bibinfo{volume}{84}},
  \bibinfo{pages}{4882} (\bibinfo{year}{2000}),
  \urlprefix\url{http://dx.doi.org/10.1103/PhysRevLett.84.4882}.

\bibitem[{\citenamefont{Borodin and Ferrari}(2009)}]{KPZ_2p1}
\bibinfo{author}{\bibfnamefont{A.}~\bibnamefont{Borodin}} \bibnamefont{and}
  \bibinfo{author}{\bibfnamefont{P.~L.} \bibnamefont{Ferrari}},
  \bibinfo{journal}{Journal of Statistical Mechanics: Theory and Experiment}
  \textbf{\bibinfo{volume}{2009}}, \bibinfo{pages}{P02009+}
  (\bibinfo{year}{2009}), ISSN \bibinfo{issn}{1742-5468},
  \urlprefix\url{http://dx.doi.org/10.1088/1742-5468/2009/02/P02009}.

\bibitem[{\citenamefont{Imamura and Sasamoto}(2012)}]{KPZ_stationary_exact}
\bibinfo{author}{\bibfnamefont{T.}~\bibnamefont{Imamura}} \bibnamefont{and}
  \bibinfo{author}{\bibfnamefont{T.}~\bibnamefont{Sasamoto}},
  \bibinfo{journal}{Physical Review Letters} \textbf{\bibinfo{volume}{108}},
  \bibinfo{pages}{190603+} (\bibinfo{year}{2012}),
  \urlprefix\url{http://dx.doi.org/10.1103/PhysRevLett.108.190603}.

\bibitem[{\citenamefont{Sasamoto and Spohn}(2010)}]{KPZ_exact}
\bibinfo{author}{\bibfnamefont{T.}~\bibnamefont{Sasamoto}} \bibnamefont{and}
  \bibinfo{author}{\bibfnamefont{H.}~\bibnamefont{Spohn}},
  \bibinfo{journal}{Physical Review Letters} \textbf{\bibinfo{volume}{104}},
  \bibinfo{pages}{230602+} (\bibinfo{year}{2010}),
  \urlprefix\url{http://dx.doi.org/10.1103/PhysRevLett.104.230602}.

\bibitem[{\citenamefont{Amir et~al.}(2011)\citenamefont{Amir, Corwin, and
  Quastel}}]{KPZ_prob_dist}
\bibinfo{author}{\bibfnamefont{G.}~\bibnamefont{Amir}},
  \bibinfo{author}{\bibfnamefont{I.}~\bibnamefont{Corwin}}, \bibnamefont{and}
  \bibinfo{author}{\bibfnamefont{J.}~\bibnamefont{Quastel}},
  \bibinfo{journal}{Communications on Pure and Applied Mathematics}
  \textbf{\bibinfo{volume}{64}}, \bibinfo{pages}{466} (\bibinfo{year}{2011}),
  ISSN \bibinfo{issn}{00103640},
  \urlprefix\url{http://dx.doi.org/10.1002/cpa.20347}.

\bibitem[{\citenamefont{Borodin et~al.}(2012)\citenamefont{Borodin, Corwin, and
  Ferrari}}]{KPZ_freeenergy_fluc}
\bibinfo{author}{\bibfnamefont{A.}~\bibnamefont{Borodin}},
  \bibinfo{author}{\bibfnamefont{I.}~\bibnamefont{Corwin}}, \bibnamefont{and}
  \bibinfo{author}{\bibfnamefont{P.}~\bibnamefont{Ferrari}}
  (\bibinfo{year}{2012}), \eprint{1204.1024},
  \urlprefix\url{http://arxiv.org/abs/1204.1024}.

\bibitem[{\citenamefont{Takeuchi and Sano}(2010)}]{KPZ_LC_PRL}
\bibinfo{author}{\bibfnamefont{K.~A.} \bibnamefont{Takeuchi}} \bibnamefont{and}
  \bibinfo{author}{\bibfnamefont{M.}~\bibnamefont{Sano}},
  \bibinfo{journal}{Physical Review Letters} \textbf{\bibinfo{volume}{104}},
  \bibinfo{pages}{230601+} (\bibinfo{year}{2010}),
  \urlprefix\url{http://dx.doi.org/10.1103/PhysRevLett.104.230601}.

\bibitem[{\citenamefont{Deegan et~al.}(1997)\citenamefont{Deegan, Bakajin,
  Dupont, Huber, Nagel, and Witten}}]{coffeering_nagel_nat}
\bibinfo{author}{\bibfnamefont{R.~D.} \bibnamefont{Deegan}},
  \bibinfo{author}{\bibfnamefont{O.}~\bibnamefont{Bakajin}},
  \bibinfo{author}{\bibfnamefont{T.~F.} \bibnamefont{Dupont}},
  \bibinfo{author}{\bibfnamefont{G.}~\bibnamefont{Huber}},
  \bibinfo{author}{\bibfnamefont{S.~R.} \bibnamefont{Nagel}}, \bibnamefont{and}
  \bibinfo{author}{\bibfnamefont{T.~A.} \bibnamefont{Witten}},
  \bibinfo{journal}{Nature} \textbf{\bibinfo{volume}{389}},
  \bibinfo{pages}{827} (\bibinfo{year}{1997}), ISSN \bibinfo{issn}{0028-0836},
  \urlprefix\url{http://dx.doi.org/10.1038/39827}.

\bibitem[{\citenamefont{Yunker et~al.}(2011)\citenamefont{Yunker, Still, Lohr,
  and Yodh}}]{coffee_ring_ellipsoids}
\bibinfo{author}{\bibfnamefont{P.~J.} \bibnamefont{Yunker}},
  \bibinfo{author}{\bibfnamefont{T.}~\bibnamefont{Still}},
  \bibinfo{author}{\bibfnamefont{M.~A.} \bibnamefont{Lohr}}, \bibnamefont{and}
  \bibinfo{author}{\bibfnamefont{A.~G.} \bibnamefont{Yodh}},
  \bibinfo{journal}{Nature} \textbf{\bibinfo{volume}{476}},
  \bibinfo{pages}{308} (\bibinfo{year}{2011}), ISSN \bibinfo{issn}{0028-0836},
  \urlprefix\url{http://dx.doi.org/10.1038/nature10344}.

\bibitem[{\citenamefont{Csahok et~al.}(1993)\citenamefont{Csahok, Honda, and
  Vicsek}}]{KPZ_quencheddisorder}
\bibinfo{author}{\bibfnamefont{Z.}~\bibnamefont{Csahok}},
  \bibinfo{author}{\bibfnamefont{K.}~\bibnamefont{Honda}}, \bibnamefont{and}
  \bibinfo{author}{\bibfnamefont{T.}~\bibnamefont{Vicsek}},
  \bibinfo{journal}{Journal of Physics A: Mathematical and General}
  \textbf{\bibinfo{volume}{26}}, \bibinfo{pages}{L171+} (\bibinfo{year}{1993}),
  ISSN \bibinfo{issn}{0305-4470},
  \urlprefix\url{http://dx.doi.org/10.1088/0305-4470/26/5/001}.

\bibitem[{\citenamefont{Sakaguchi}(2010)}]{KPZ_quencheddis_SOC}
\bibinfo{author}{\bibfnamefont{H.}~\bibnamefont{Sakaguchi}},
  \bibinfo{journal}{Physical Review E} \textbf{\bibinfo{volume}{82}},
  \bibinfo{pages}{032101+} (\bibinfo{year}{2010}),
  \urlprefix\url{http://dx.doi.org/10.1103/PhysRevE.82.032101}.

\bibitem[{\citenamefont{Sneppen}(1992)}]{KPZ_quencheddis_pinning}
\bibinfo{author}{\bibfnamefont{K.}~\bibnamefont{Sneppen}},
  \bibinfo{journal}{Physical Review Letters} \textbf{\bibinfo{volume}{69}},
  \bibinfo{pages}{3539} (\bibinfo{year}{1992}),
  \urlprefix\url{http://dx.doi.org/10.1103/PhysRevLett.69.3539}.

\bibitem[{\citenamefont{Champion et~al.}(2007)\citenamefont{Champion, Katare,
  and Mitragotri}}]{ellipsoidstretch_PNAS}
\bibinfo{author}{\bibfnamefont{J.~A.} \bibnamefont{Champion}},
  \bibinfo{author}{\bibfnamefont{Y.~K.} \bibnamefont{Katare}},
  \bibnamefont{and}
  \bibinfo{author}{\bibfnamefont{S.}~\bibnamefont{Mitragotri}},
  \bibinfo{journal}{Proceedings of the National Academy of Sciences}
  \textbf{\bibinfo{volume}{104}}, \bibinfo{pages}{11901}
  (\bibinfo{year}{2007}),
  \urlprefix\url{http://dx.doi.org/10.1073/pnas.0705326104}.

\bibitem[{\citenamefont{Ho}(1993)}]{ellipsoid_stretch_first}
\bibinfo{author}{\bibnamefont{Ho}}, \bibinfo{journal}{Colloid and Polymer
  Science} \textbf{\bibinfo{volume}{271}}, \bibinfo{pages}{469}
  (\bibinfo{year}{1993}).

\bibitem[{\citenamefont{Loudet et~al.}(2006)\citenamefont{Loudet, Yodh, and
  Pouligny}}]{ellipsoid_wetting_yodh}
\bibinfo{author}{\bibfnamefont{J.~C.} \bibnamefont{Loudet}},
  \bibinfo{author}{\bibfnamefont{A.~G.} \bibnamefont{Yodh}}, \bibnamefont{and}
  \bibinfo{author}{\bibfnamefont{B.}~\bibnamefont{Pouligny}},
  \bibinfo{journal}{Physical Review Letters} \textbf{\bibinfo{volume}{97}},
  \bibinfo{pages}{018304+} (\bibinfo{year}{2006}),
  \urlprefix\url{http://dx.doi.org/10.1103/PhysRevLett.97.018304}.

\bibitem[{\citenamefont{Loudet et~al.}(2005)\citenamefont{Loudet, Alsayed,
  Zhang, and Yodh}}]{capillary_interactions_ellipsoids_yodh}
\bibinfo{author}{\bibfnamefont{J.~C.} \bibnamefont{Loudet}},
  \bibinfo{author}{\bibfnamefont{A.~M.} \bibnamefont{Alsayed}},
  \bibinfo{author}{\bibfnamefont{J.}~\bibnamefont{Zhang}}, \bibnamefont{and}
  \bibinfo{author}{\bibfnamefont{A.~G.} \bibnamefont{Yodh}},
  \bibinfo{journal}{Physical Review Letters} \textbf{\bibinfo{volume}{94}},
  \bibinfo{pages}{018301+} (\bibinfo{year}{2005}),
  \urlprefix\url{http://dx.doi.org/10.1103/PhysRevLett.94.018301}.

\bibitem[{\citenamefont{Bowden et~al.}(2001)\citenamefont{Bowden, Arias, Deng,
  and Whitesides}}]{interface_attract_plates}
\bibinfo{author}{\bibfnamefont{N.}~\bibnamefont{Bowden}},
  \bibinfo{author}{\bibfnamefont{F.}~\bibnamefont{Arias}},
  \bibinfo{author}{\bibfnamefont{T.}~\bibnamefont{Deng}}, \bibnamefont{and}
  \bibinfo{author}{\bibfnamefont{G.~M.} \bibnamefont{Whitesides}},
  \bibinfo{journal}{Langmuir} \textbf{\bibinfo{volume}{17}},
  \bibinfo{pages}{1757} (\bibinfo{year}{2001}),
  \urlprefix\url{http://dx.doi.org/10.1021/la001447o}.

\bibitem[{\citenamefont{Brown et~al.}(2000)\citenamefont{Brown, Smith, and
  Rennie}}]{interface_anisotropicparticles}
\bibinfo{author}{\bibfnamefont{A.~B.~D.} \bibnamefont{Brown}},
  \bibinfo{author}{\bibfnamefont{C.~G.} \bibnamefont{Smith}}, \bibnamefont{and}
  \bibinfo{author}{\bibfnamefont{A.~R.} \bibnamefont{Rennie}},
  \bibinfo{journal}{Physical Review E} \textbf{\bibinfo{volume}{62}},
  \bibinfo{pages}{951} (\bibinfo{year}{2000}),
  \urlprefix\url{http://dx.doi.org/10.1103/PhysRevE.62.951}.

\bibitem[{\citenamefont{Madivala et~al.}(2009)\citenamefont{Madivala, Fransaer,
  and Vermant}}]{ellipsoids_interface_rheology}
\bibinfo{author}{\bibfnamefont{B.}~\bibnamefont{Madivala}},
  \bibinfo{author}{\bibfnamefont{J.}~\bibnamefont{Fransaer}}, \bibnamefont{and}
  \bibinfo{author}{\bibfnamefont{J.}~\bibnamefont{Vermant}},
  \bibinfo{journal}{Langmuir} \textbf{\bibinfo{volume}{25}},
  \bibinfo{pages}{2718} (\bibinfo{year}{2009}),
  \urlprefix\url{http://dx.doi.org/10.1021/la803554u}.

\bibitem[{\citenamefont{Vandebril et~al.}(2009)\citenamefont{Vandebril,
  Fransaer, and Vermant}}]{emulsions_ellipsoids_softmatter}
\bibinfo{author}{\bibfnamefont{S.}~\bibnamefont{Vandebril}},
  \bibinfo{author}{\bibfnamefont{J.}~\bibnamefont{Fransaer}}, \bibnamefont{and}
  \bibinfo{author}{\bibfnamefont{J.}~\bibnamefont{Vermant}},
  \bibinfo{journal}{Soft Matter} \textbf{\bibinfo{volume}{5}},
  \bibinfo{pages}{1717} (\bibinfo{year}{2009}),
  \urlprefix\url{http://dx.doi.org/10.1039/b816680c}.

\bibitem[{\citenamefont{Park and Furst}(2011)}]{interface_att_furst}
\bibinfo{author}{\bibfnamefont{B.~J.} \bibnamefont{Park}} \bibnamefont{and}
  \bibinfo{author}{\bibfnamefont{E.~M.} \bibnamefont{Furst}},
  \bibinfo{journal}{Soft Matter} \textbf{\bibinfo{volume}{7}},
  \bibinfo{pages}{7676} (\bibinfo{year}{2011}),
  \urlprefix\url{http://dx.doi.org/10.1039/c1sm00005e}.

\bibitem[{SOM()}]{SOM_}
\bibinfo{journal}{See accompanying Supplemental Information document.}  (????).

\bibitem[{\citenamefont{Leschhorn}(1996)}]{KPZ_qd_anisotropicdepinning}
\bibinfo{author}{\bibfnamefont{H.}~\bibnamefont{Leschhorn}},
  \bibinfo{journal}{Physical Review E} \textbf{\bibinfo{volume}{54}},
  \bibinfo{pages}{1313} (\bibinfo{year}{1996}),
  \urlprefix\url{http://dx.doi.org/10.1103/PhysRevE.54.1313}.

\bibitem[{\citenamefont{Merton}(1968)}]{mattheweffect}
\bibinfo{author}{\bibfnamefont{R.~K.} \bibnamefont{Merton}},
  \bibinfo{journal}{Science} \textbf{\bibinfo{volume}{159}},
  \bibinfo{pages}{56} (\bibinfo{year}{1968}), ISSN \bibinfo{issn}{1095-9203},
  \urlprefix\url{http://dx.doi.org/10.1126/science.159.3810.56}.

\bibitem[{\citenamefont{Dickinson}(2010)}]{food_emulsions}
\bibinfo{author}{\bibfnamefont{E.}~\bibnamefont{Dickinson}},
  \bibinfo{journal}{Current Opinion in Colloid \& Interface Science}
  \textbf{\bibinfo{volume}{15}}, \bibinfo{pages}{40} (\bibinfo{year}{2010}),
  ISSN \bibinfo{issn}{13590294},
  \urlprefix\url{http://dx.doi.org/10.1016/j.cocis.2009.11.001}.

\bibitem[{\citenamefont{Vermant}(2011)}]{coffee_ring_ellipsoids_newsviews}
\bibinfo{author}{\bibfnamefont{J.}~\bibnamefont{Vermant}},
  \bibinfo{journal}{Nature} \textbf{\bibinfo{volume}{476}},
  \bibinfo{pages}{286} (\bibinfo{year}{2011}), ISSN \bibinfo{issn}{0028-0836},
  \urlprefix\url{http://dx.doi.org/10.1038/476286a}.

\end{thebibliography}

\end{document}